\documentclass[jcp,10pt,preprint]{revtex4-1}

\usepackage{amsmath}    
\usepackage{graphicx}   
\usepackage{verbatim}   
\usepackage{color}      
\usepackage{subfigure}  
\usepackage{hyperref}   
\usepackage{multirow}

\usepackage{amsfonts}
\begin{document}
\title{Bifurcation of Transition Paths Induced by Coupled Bistable Systems}
\author{Chengzhe Tian}
\email{chengzhe@nbi.dk}
\affiliation{Niels Bohr Institute, University of Copenhagen, Blegdamsvej 17, 2100 Copenhagen, Denmark}
\author{Namiko Mitarai}
\email{mitarai@nbi.dk}
\affiliation{Niels Bohr Institute, University of Copenhagen, Blegdamsvej 17, 2100 Copenhagen, Denmark}
\date{\today}

\begin{abstract}

We discuss the transition paths in a coupled bistable system consisting of interacting multiple identical bistable motifs. We propose a simple model of coupled bistable gene circuits as an example, and show that its transition paths are bifurcating. We then derive a criterion to predict the bifurcation of transition paths in a generalized coupled bistable system. We confirm the validity of the theory for the example system by numerical simulation. We also demonstrate in the example system that, if the steady states of individual gene circuits are not changed by the coupling, the bifurcation pattern is not dependent on the number of gene circuits.  We further show that the transition rate exponentially decreases with the number of gene circuits when the transition path does not bifurcate, while a bifurcation facilitates the transition by lowering the quasi-potential energy barrier.

\end{abstract}

\maketitle

\section{Introduction}

Bistable systems are widely utilized to model the biological processes which exhibit distinct phenotypes under homogeneous conditions \cite{Ferrell2001, Ferrell2002, Veening2008}. Switching between phenotypes (stable states) is facilitated by the stochasticity arising from molecular noise \cite{Balazsi2011}. The paths of switching have been studied in various systems, such as the $\lambda$-phage lysis-lysogeny decision \cite{Aurell2002} and cellular development and differentiation \cite{Wang2011}, to gain insights into the molecular processes of biological decision making.

In nature, we sometimes observe situations where multiple bistable systems are coupled. For example in bacterial quorum sensing, every cell produces small signaling molecules whose production is regulated by a positive feedback to synchronize the population. This positive feedback may induce bistability between the high and low concentrations of the signaling molecules. The cells are further coupled by secreting and sensing the signaling molecules in the medium\cite{Miller2001,Muller2006}. Another example is the Toxin-Antitoxin (TA) loci in {\em Escherichia coli}, where there are 10 known mRNase toxins and every pair may act as a bistable system allowing the cells to switch between the normal growing state and the dormant state which exhibits antibiotic persistence \cite{Maisonneuve2014,Cataudella2012, Fasani2013}. The TA systems may interact each other by interfering the protein synthesis and the cellular growth.

Motivated by these systems, in this paper we analyze a coupled bistable system that consists of interacting identical bistable motifs. We consider the case where coupling is such that the coupled system itself is also bistable and for each stable state the individual motifs are in the same steady state. In other word, the coupling is positive to allow all the bistable motifs to jointly switch from one state to the other state. While the individual bistable systems without coupling show the same pattern of transition paths by definition, the switching properties of the coupled system remain unclear. The individual systems may transit synchronously at the same pace, resulting in one transition path. However, it is also possible that individual systems lead the switching process. As a result, the transition paths of the coupled system are split into multiple ones, a phenomenon called "bifurcation of transition path". Since many properties of the coupled system, such as the transition rates between the steady states, are dependent on the transition paths, it is interesting to study whether the transition paths of a given coupled bistable system bifurcate and how this bifurcation relates to the individual bistable systems and their coupling. 

Bifurcation of transition paths was first demonstrated in the Maier-Stein model \cite{Maier1996}, but its relevance and implication to biochemical systems remain to be explored. In this work, we first construct a model of coupled bistable gene circuits in Section \ref{modeling}. Then we demonstrate that the transition paths bifurcate with appropriate parameter sets. In Section \ref{theory}, we construct a general formulation of coupled bistable systems. We consider the transition of this general model between its steady states as a noise-induced exit process from a metastable state and propose a criterion for the bifurcation of transition paths by extending the previous works on the Maier-Stein model \cite{Maier1996}. Finally in Section \ref{application}, we apply our criterion to the model of coupled bistable gene circuits. We confirm the theory numerically and discuss the transition rates. 

\section{Model of Coupled Bistable Gene Circuits}\label{modeling}

\begin{figure}
  \centering\includegraphics{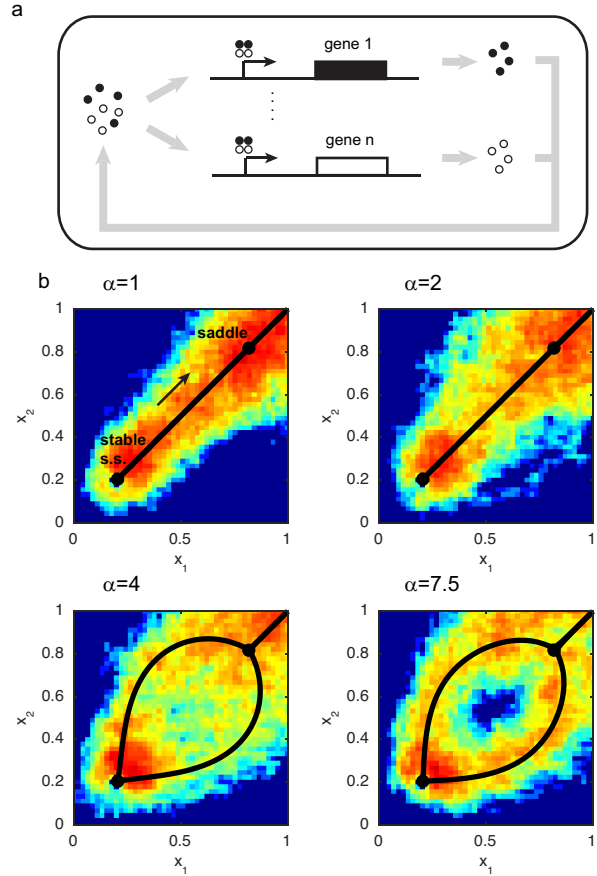}
  \caption{\label{model}(Color online) Model of coupled bistable gene circuits.
    {\bf a}. A schematic illustration of the model. A cell contains $n$ genes with weak promoters and their products mix in the cell and activate the gene expression by binding to the promoters in the form of tetramers.
    {\bf b}. Distribution of transition paths from the lower stable steady state ($\mathbf{x}_l$) to the higher one ($\mathbf{x}_h$). We set $V=45$ and we carry out 100 simulations using Gillespie algorithm. The distribution of the last instanton trajectories, i.e. the trajectories associated with the successful escapes, is calculated. We present the frequencies in the logarithmic scale in the form of heat plots. Red indicates high frequency and blue indicates low frequency. The black lines represent the most probable escape paths computed in the zero-noise limit. The parameter values are $k_0=0.1$, $k_1=1$, $S=1$ and $\gamma=0.5$. In Appendix D, we verify that the Gillespie simulation is carried out in the low-noise limit.
    }
\end{figure}

Consider a model of coupled bistable gene circuits (Fig. \ref{model}a). First we restrict our attention to one gene. The promoter of the gene is weak and the proteins of this gene bind to the promoter in the form of tetramers to activate the gene expression. We may model the proteins of this gene using
\begin{equation}\label{example1d}
  \frac{\mathrm{d}x}{\mathrm{d}t} = k_0 + k_1 \frac{x^4}{x^4 + S^4} - \gamma x
\end{equation}
where $x$ is the concentration of the protein, $k_0$ refers to the basal synthesis rate of the protein, the Hill term describes the activation of gene expression by the tetramers and the last term models the linear degradation. With appropriate parameter values, the positive feedback on gene expression allows Eq. \ref{example1d} to show bistability. 

We now couple $n$ such genes in one cell and we assume that all these genes (and their promoters and proteins) have the same kinetic properties. The genes are coupled in a way that the proteins of the genes are well-mixed in the cells and the mixture activates a cell by binding to a promoter in a tetramer. Multiple coupling strategies may be used. For example, if the proteins of the $n$ genes are identical, any four monomers may bind a promoter and we may model the coupled bistable gene circuits using 
\begin{equation}\label{alpha1}
  \frac{\mathrm{d}x_i}{\mathrm{d}t} = k_0 + k_1\frac{(x_1+\cdots+x_n)^4}{(x_1+\cdots+x_n)^4 + S_1^4} - \gamma x_i
\end{equation}
where $x_i$ is the concentration of the products of the $i$-th gene and $S_1$ is the Hill constant for the coupled system. Meanwhile, if the genes are equipped with identical promoters but encode different proteins, and the tetramer activating the gene expression consists of four monomers from the same gene, we model the coupled system using
\begin{equation}\label{alpha4}
  \frac{\mathrm{d}x_i}{\mathrm{d}t} = k_0 + k_1\frac{x_1^4+\cdots+x_n^4}{x_1^4+\cdots+x_n^4 + S_2^4} - \gamma x_i
\end{equation}

We then generalize these examples and we propose the following model of coupled bistable gene circuits
\begin{equation}\label{example}
  \frac{\mathrm{d}x_i}{\mathrm{d}t} = k_0 + k_1 \frac{ \bar{x}^4 }{ \bar{x}^4 + S^4 } - \gamma x_i, \bar{x}=\left(\frac{1}{n}\sum_{i=1}^n x_i^\alpha \right)^{1/\alpha}
\end{equation}
where the parameter $\alpha$, called the configuration parameter, governs the general coupling strategy. It is straightforward that $\alpha=1$ corresponds to Eq. \ref{alpha1} and $\alpha=4$ corresponds to Eq. \ref{alpha4}. Here we allow $\alpha$ to be arbitrary positive values, though not all coupling strategies are biologically plausible. We also choose the values of the Hill constants ($S_1$ in Eq. \ref{alpha1} and $S_2$ in Eq. \ref{alpha4}) such that the steady states of every individual gene are not affected by the coupling. If the model for one gene (Eq. \ref{example1d}) is bistable, it is straightforward that the general model of coupled bistable gene circuits (Eq. \ref{example}) is also bistable and contains three steady states: two stable ones and one saddle. Furthermore, one can show that $x_1=x_2=\cdots=x_n$ holds at every steady state. For convenience, throughout this work we call the stable state where the concentrations of all gene products are low the "lower stable steady state" $\mathbf{x}_l$ and the other stable steady state the "higher stable steady state" $\mathbf{x}_h$. 

If we set the volume of the cell to be $V$, the concentrations of proteins $x_i$ can be converted to the absolute numbers of molecules ($=Vx_i$, should be an integer) and the coupled system is governed by the chemical master equation. One may then sample the transition paths between the two steady states using the Gillespie algorithm \cite{Gillespie1977}. Here we restrict our attention to $n=2$ and the transition from the lower stable steady state to the higher one. As illustrated in Fig. \ref{model}b, when $\alpha=1$, the transition paths are narrowly distributed around the diagonal (the line satisfying $x_1=x_2$), suggesting that the switching of the two genes is synchronized. The distribution becomes wider as the value of $\alpha$ increases, and at $\alpha=7.5$, the transition paths exhibit a visible bifurcation. Therefore, coupled bistable systems are capable of exhibiting both bifurcating and non-bifurcating transition paths, and we can modulate the bifurcation pattern for the model of coupled bistable gene circuits with the parameter $\alpha$.

\section{Theory}\label{theory}

In this section we develop a criterion for coupled bistable systems to predict whether the transition paths bifurcate or not. To formulate a general model for coupled bistable systems, we notice that in the model of coupled bistable gene circuits (Eq. \ref{example}), every gene is governed by the concentration of its own protein ($x_i$, in the degradation term) and the average concentration of all proteins ($\bar{x}$, in the production term). Here for a general coupled bistable system, we may model the deterministic drifts of the individual bistable systems in the same fashion and describe the effect of noises using
\begin{equation}\label{model_defn}
  \frac{\mathrm{d}x_i}{\mathrm{d}t} = f(x_i, h(\mathbf{x})) + \sqrt{\epsilon}\sqrt{g(x_i, h(\mathbf{x}))} \xi_i, i=1,2,\cdots,n
\end{equation}
which is interpreted as an Ito-type stochastic differential equation. Here $x_i$ is the state of the $i$-th bistable system and $n$ is the number of systems to be coupled. The function $h$, defined as $h(\mathbf{x})=(\sum_{i=1}^n x_i^\alpha/n)^{1/\alpha}$, computes the average state. We choose this formulation because it allows modulating of $\alpha$ and $n$ without changing the steady states. Obviously, the state of every individual bistable system is governed by its own state and the average state of all bistable systems, as modeled by the deterministic drift $f$. The second term of Eq. \ref{model_defn} arises from expanding the chemical master equation in the continuous limit and serves as the noise term for the coupled bistable systems. Here $\xi_i$ are independent Gaussian white noise sources ($\langle \xi_i(t)\xi_j(t')\rangle = \delta_{i,j}\delta(t-t')$) and we assume the multiplicative noise $g(x_i, h(\mathbf{x}))$ for the $i$-th system is also a function of its own state and the average state. The overall noise is modulated by a small parameter $0<\epsilon\ll 1$ in order to keep the validity of the continuous limit. We assume that the deterministic drift with $n=1$ gives three steady states - two stable ones and an unstable one. We further assume the coupled system also allows three steady states - two stable ones and a saddle point. At each steady state, the states of all bistable systems are assumed to be equal. A wide range of coupled bistable systems may be modeled in the fashion of Eq. \ref{model_defn}. For example, the model of coupled bistable system, as mentioned in Section \ref{modeling}, corresponds to 
\begin{equation}\label{fuv}
  f(u,v) = k_0 + k_1 v^4/(v^4+S^4) - \gamma u
\end{equation} and a noise amplitude of the form 
\begin{equation}\label{guv}
  g(u,v) = k_0 + k_1 v^4/(v^4+S^4) + \gamma u
\end{equation}
\cite{Note}. The parameter $\alpha$ in $h(\mathbf{x})$ corresponds to the coupling mode of the genes and the small $\epsilon$ corresponds to a cell with a large volume $V$ (in the large volume limit, $\epsilon \propto V^{-1}$). 

We derive our criterion by performing linear perturbation analysis in the zero-noise limit. When the noise level of a system reaches zero ($\epsilon\rightarrow 0$), the transition between stable steady states is dominated by the path(s) associated with the lowest energy cost (or "action" in standard terminology). This path is called the Most Probable Escape Path (MPEP)\cite{Maier1993}. The MPEP can be quantified analytically since it is governed by the Freidlin-Wentzell Hamiltonian \cite{Freidlin2012}, allowing ones to use the tools of analytical mechanics for derivation. 

One property of MPEP in conventional cases is that the MPEP of transition passes through the saddle of the system (refer to \cite{Maier1997}) and after that the system follows the deterministic flow to the final stable steady state \cite{Maier1993}. In our formulation of coupled bistable systems Eq. \ref{model_defn}, the individual bistable systems are always driven by the same deterministic drift after passing the saddle. We then conclude that no bifurcation of transition paths will occur between saddle and the final steady state. Therefore, the transition capable of bifurcation is the one from the initial stable steady state to the saddle, and we may restrict our attention to this exit process.

The exit process may be studied with the approach of linear perturbation. To be precise, we first assume that the MPEP does not bifurcate, i.e. all the bistable systems take the same state during the transition. We then perturb the shape of MPEP by allowing some bistable systems to be in different states from the others during transition and we examine how the associated actions for the perturbed MPEP change. If the actions always increase regardless of the perturbation, the non-bifurcating path is locally energetically favorable. For the model of coupled bistable gene circuits, we confirm by numerical simulation that this path is indeed the MPEP. Meanwhile, if any perturbation leads to a decreased action, the non-bifurcating path is energetically unfavorable. The MPEP should then be some other paths. Due to the symmetry of the coupled system, multiple paths must exist. Therefore, the coupled system exhibits a bifurcation of transition paths.

We now follow this approach and we begin our analysis by deriving the equation governing the MPEP, closely following the procedure found in previous work\cite{Maier1996, Maier1997}. The model of the coupled bistable system Eq. \ref{model_defn}, viewed as a stochastic differential equation, gives the Fokker-Planck equation \cite{Risken1996}
\begin{equation}\label{Fokker-Planck}
  \frac{\partial}{\partial t} P(\mathbf{x},t) = -\nabla\cdot (\mathbf{F}P(\mathbf{x},t)) + \frac{\epsilon}{2} \nabla\cdot[\nabla\cdot (\mathbf{G}P(\mathbf{x},t))]
\end{equation}
where $P(\mathbf{x},t)$ denotes the probability of the coupled system at state $\mathbf{x}$ and time $t$. The drift vector $\mathbf{F}=[F_1,F_2,\cdots,F_n]'$ satisfies $F_i=f(x_i,h(\mathbf{x}))$ and the covariance $\mathbf{G}$ is a diagonal matrix satisfying $G_{i,i} = g(x_i, h(\mathbf{x}))$. As the initial condition, we use a delta function at the initial stable steady state. The separatrix of the basins gives an absorbing boundary condition. In the zero-noise limit, $\partial P(\mathbf{x},t)/\partial t\approx 0$ and we replace $P(\mathbf{x},t)$ by the quasi-steady state distribution $P_{ss}(\mathbf{x})$ governed by \cite{Naeh1990}
\begin{equation}\label{FP_quasi}
  -\nabla\cdot (\mathbf{F}P_{ss}(\mathbf{x})) + \frac{\epsilon}{2} \nabla\cdot[\nabla\cdot (\mathbf{G}P_{ss}(\mathbf{x}))] = 0
\end{equation}
We further assume the quasi-steady state distribution to take an Arrhenius form $P_{ss}(\mathbf{x}) \propto \exp\{-W(\mathbf{x})/\epsilon\}$, where $W(\mathbf{x})$ is the quasi-potential\cite{Kupferman1992}. We perform a WKB expansion by plugging the form into Eq. \ref{FP_quasi} and keeping the terms of the lowest order of $\epsilon$. One can show that the MPEP of the transition is a classical zero-energy trajectory of the Freidlin-Wentzell Hamiltonian 
\begin{equation}\label{FW_Hamiltonian}
  \mathcal{H}(\mathbf{x},\mathbf{p})=\frac{1}{2}\mathbf{p}^T \mathbf{G}(\mathbf{x}) \mathbf{p} + \mathbf{F}(\mathbf{x})^T\mathbf{p},
\end{equation}
where the momentum vector can be computed by $\mathbf{p} = \nabla W(\mathbf{x})$\cite{Maier1997,Note3}. The equation $\mathcal{H}=0$ can be viewed as the equation governing the MPEP, but it is worth noting that not all trajectories satisfying $\mathcal{H}=0$ are the MPEPs, as one has to examine the associated actions.

We now determine whether the MPEP of the transition bifurcates or not by linearly perturbing the non-bifurcating path and examining the actions. We can show that the non-bifurcating path is a zero-energy trajectory of the Freidlin-Wentzell Hamiltonian (Appendix A). The action associated with the non-bifurcating path $\bar{\mathbf{x}}=[\bar x, \bar x, \cdots, \bar x]^T$ can then be computed by the quasi-potential $W(\mathbf{x})$ and one may quantify how the actions change with perturbations to this path by considering the expansion $W(\bar{\mathbf{x}}+\Delta\mathbf{x}) = W(\bar{\mathbf{x}}) + \nabla W(\bar{\mathbf{x}})\cdot\Delta\mathbf{x} + \frac{1}{2} \Delta\mathbf{x}^T\nabla\nabla W(\bar{\mathbf{x}})\Delta\mathbf{x}$. Due to symmetry, the first-order term vanishes if the system is perturbed in the directions perpendicular to the non-bifurcating path. Note that these directions are of our primary interest, and we then discuss the second-order term and examine the eigenvalues of the hessian matrix $\mathbf{Z(\bar{\mathbf{x}})}=\nabla\nabla W(\bar{\mathbf{x}})$. The equation governing the hessian matrix, computed by differentiating the $\mathcal{H}=0$ twice over $\mathbf{x}$ and making use of the Hamiltonian equation $\mathrm{d}x_i/\mathrm{d}t = \partial\mathcal{H}/\partial p_i$ (also refer to Ref. \cite{Maier1996}), is
\begin{align}
  \frac{\mathrm{d}\mathbf{Z}}{\mathrm{d}t} =& -\mathbf{Z}\mathbf{G}\mathbf{Z} - \mathbf{B}^T\mathbf{Z} - \mathbf{Z}\mathbf{B} - \sum_k p_k\nabla\nabla F_k \notag\\
     &- \mathbf{C}^T\mathbf{Z} - \mathbf{Z}\mathbf{C} - \frac{1}{2}\sum_k p_k^2\nabla\nabla G_{k,k} \label{hessian_full}
\end{align}
where $B_{i,j} = \partial F_i/\partial x_j$ is a linearization of the drift vector. The linearization of the variances gives $C_{i,j} = p_i\partial G_{ii}/\partial x_j$. 

We may simplify Eq. \ref{hessian_full} significantly (details in Appendix B). By utilizing the fact that all the bistable systems share the same state along the non-bifurcating path, we may express the hessian matrix $\mathbf{Z} = z_1\mathbf{I}_{n\times n}+z_2\mathbf{1}_{n\times n}$, where $\mathbf{I}_{n\times n}$ is an $n\times n$ identity matrix and $\mathbf{1}_{n\times n}$ is an $n\times n$ one matrix. We can further express the remaining terms, namely $\mathbf{B}$, $\mathbf{C}$, $\nabla\nabla F_k$ and $\nabla\nabla G_{k,k}$, in a similar fashion. By plugging the new forms into Eq. \ref{hessian_full}, we obtain the equations governing $z_1$ and $z_2$: $z_1$ is shown to follow
\begin{align}
  \frac{\mathrm{d}z_1}{\mathrm{d}t} =& -2z_1g_1'p - g z_1^2 - \frac{1}{2}p^2\left(g_{11}''+g_2'\frac{\alpha-1}{x}\right) \notag\\
    & - p\left(f_{11}''+f_2'\frac{\alpha-1}{x}\right) - 2z_1f_1'\label{z1_dt}
\end{align}
and $z_2$ is shown to follow Eq. \ref{z2_dt}. Here $f_1'(u,v) = \partial f(u,v)/\partial u$, $f_2'(u,v)=\partial f(u,v)/\partial v$ and $f_{11}''(u,v) = \partial^2 f(u,v)/\partial u^2$ and similar for $g$. In Eq. \ref{z1_dt}, $x$ and $p$ refer to the state and the momentum of every bistable system. The two arguments to the function $f$ and $g$ are $x$. 

Here we analyze the eigenvalues of the hessian matrix $\mathbf{Z}$, namely $z_1$ (repeat $n-1$ times) and $z_1+nz_2$, and we need to determine the eigenvalues corresponding to perturbations that break the non-bifurcating assumption. In the Appendix C we show that the eigenvalue $z_1+nz_2$ is only governed by one individual bistable system and it represents the change in action along the non-bifurcating path. So this eigenvalue is not relevant in our context and we focus on the directions perpendicular to the non-bifurcating path, which is given by the eigenvalue $z_1$. We should examine the sign of $z_1$ along the non-bifurcating path from the initial stable steady state to the saddle. If $z_1$ is always positive, the perturbation is energetically unfavorable and the non-perturbing path is locally stable. If $z_1$ is negative somewhere, the MPEP bifurcates into multiple paths. 

Sometimes it is convenient to express $z_1$ as a function of $x$ rather than time. Based on the Freidlin-Wentzell hamiltonian for the coupled bistable system, we have that the momentum vector may be computed by $\mathbf{p}=\mathbf{G}^{-1}(\dot{\mathbf{x}} - \mathbf{F})$. In addition, the transition for the individual bistable system satisfies that $\mathrm{d}x/\mathrm{d}t=-f(x,x)$ since it is moving against the deterministic drift \cite{Maier1996}. By plugging these relations into Eq. \ref{z1_dt}, we show that 
\begin{align}
  \frac{\mathrm{d}z_1}{\mathrm{d}x} =& \frac{2f}{g^2}\left(g_{11}''+g_2'\frac{\alpha-1}{x}\right) - \frac{2}{g}\left(f_{11}''+f_2'\frac{\alpha-1}{x}\right) \notag\\
    & -\frac{4z_1g_1'}{g} + \frac{g}{f}z_1^2  + 2\frac{z_1f_1'}{f} \label{z1}
\end{align}
Usually we assume that the system is equipped with a constant, non-zero noise at the initial stable steady states \cite{Maier1996}. We may then solve the equation $\mathrm{d}z_1/\mathrm{d}x=0$ at the initial stable steady states and we set the nontrivial solution to be the initial condition, which is $z_1=-2f_1'/g$. We can integrate Eq. \ref{z1} from the initial stable steady state to the saddle and determine whether the MPEP bifurcates or not by looking at the sign of $z_1$. 

Eq. \ref{z1} generates several insights. The first two terms of Eq. \ref{z1} illustrate that the nonlinearity induces the bifurcation of MPEP: $f_{11}''$ and $g_{11}''$ gives information about the nonlinearity for a bistable system without coupling, while $\alpha\neq 1$ gives the nonlinearity created by the coupling of the bistable system. We also notice that Eq. \ref{z1} contains no terms of $n$, indicating that the bifurcation of MPEP is not dependent on the number of bistable systems to be coupled.

\section{Application}\label{application}

\begin{figure}
  \centering\includegraphics{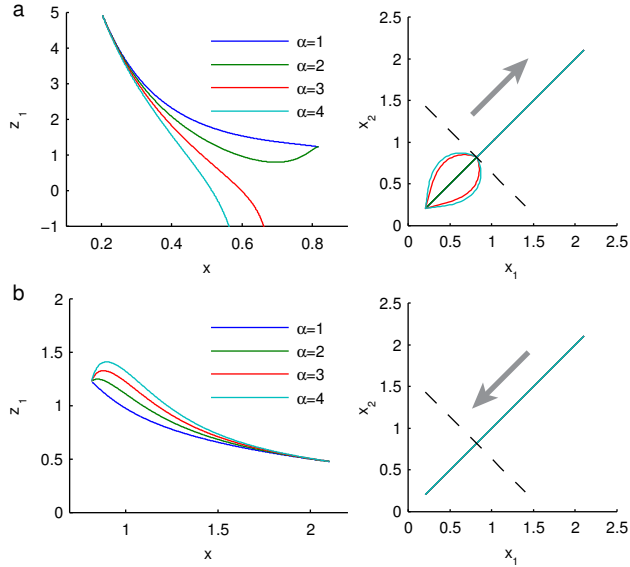}
  \caption{\label{2d_result}(Color online) Analysis of the coupling of two genes. 
    The eigenvalues induced by the coupling of genes were integrated along the path with no bifurcation from the initial stable steady state to the saddle and plotted in the left panels. The right panels illustrate the most probable escape paths which are numerically computed with gMAM method. The dashed lines represent the separatrix of the basins surrounding the two stable steady states. The arrows represent the directions of transition.
    {\bf a}. Forward transition from the lower stable steady state ($\mathbf{x}_l$) to the higher one ($\mathbf{x}_h$).
    {\bf b}. Backward transition from the higher stable steady state ($\mathbf{x}_h$) to the lower one ($\mathbf{x}_l$).
  }
\end{figure}

In this section we apply our theoretical criterion to the model of coupled bistable gene circuits (Eq. \ref{fuv}-\ref{guv}), and we first consider the MPEP of the model with two coupled genes. For the forward transition from the lower stable steady state $\mathbf{x}_l$ to the higher one $\mathbf{x}_h$, if we choose $\alpha=1$, i.e. no nonlinearity is generated from the coupling of the genes, and integrate Eq. \ref{z1} from $\mathbf{x}_l$ to the saddle, we find that the eigenvalue $z_1$ monotonously decreases along the path (Fig. \ref{2d_result}a, left panel). $z_1$ remains positive throughout the path, indicating that at $\alpha=1$ the MPEP does not bifurcate. For $\alpha=2$, the nonlinearity from the coupling of the genes drives $z_1$ to a lower value, but $z_1$ remains positive and the MPEP does not bifurcate as well. For $\alpha=3$ and 4, the coupling is highly nonlinear and $z_1$ reaches negative values and diverges, suggesting that the MPEP bifurcates. Meanwhile, for the backward transition from the higher stable steady state $\mathbf{x}_h$ to the lower one $\mathbf{x}_l$, we may integrate Eq. \ref{z1} from $\mathbf{x}_h$ to the saddle and we find that the eigenvalue $z_1$ remains positive regardless of the values of $\alpha$ (Fig. \ref{2d_result}b, left panel), suggesting that the MPEP never bifurcates. 

To verify our prediction, we use the geometric Minimum Action Method (gMAM) \cite{Heymann2008PRL,Heymann2008CPAM} to compute the MPEP numerically (Fig. \ref{2d_result}ab, right panels)\cite{Note2}. As predicted, the MPEP for the forward transition with $\alpha=1$ and $2$ do not bifurcate, the one with $\alpha=3$ and $4$ bifurcate to two symmetric ones with equal actions and no bifurcation can be observed for the backward transition. Therefore, the predictive power of the theoretical criterion is confirmed and this bifurcation pattern can explain the distribution of transition paths illustrated in Fig. \ref{model}b.

\begin{figure}
  \centering\includegraphics{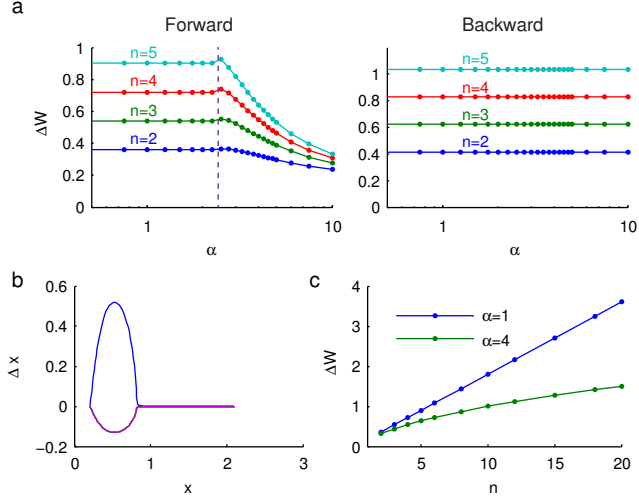}
  \caption{\label{nd_result}(Color online) Analysis of the coupling of multiple genes.
    {\bf a}. The actions $\Delta W$ (defined as the difference in the quasi-potential between the initial stable steady state and the saddle) are computed numerically for both forward transitions (left panels) and backward transitions (right panels). We consider the coupling of two to five genes and sample the configuration $\alpha$ in a wide range (the sampled values are shown in dots). The dash line represents the predicted critical $\alpha$ for bifurcation. The theory predicts that the transition paths for systems with $\alpha\approx2.5$ are bifurcated and the action becomes smaller. With the present numerical accuracy, we are able to show that the transition paths are bifurcated, but not able to find the correct action.
    {\bf b}. The most probable escape path of the forward transition for systems with $n=5$ and $\alpha=4$. The horizontal axis represents the arithmetic average of the genes' states ($x = \sum_{i=1}^n x_i/n$) and the vertical axis represents the difference between the state of individual genes and the average state ($\Delta x_i=x_i-x$).
    {\bf c}. The actions $\Delta W$ of forward transitions for multi-dimensional systems with $\alpha=1$ (no bifurcation) and 4 (bifurcation).
  }
\end{figure}

We then examine the transition rates of two coupled genes. The action associated with the MPEP ($\Delta W$) is defined as the difference in the quasi-potential between the initial and final stable steady states, and one can show that the transition rate is proportional to $\exp\{-\Delta W/\epsilon\}$ \cite{Maier1997}. We sample the configuration $\alpha$ in a wide range and compute the actions. In accordance with the Maier-Stein model \cite{Maier1996}, the actions for the model of coupled bistable gene circuits are not modulated by the value of $\alpha$ if no bifurcation of MPEP exists (Fig. \ref{nd_result}a, $n=2$). Meanwhile, in the presence of bifurcation, the actions are driven to a lower value by this configuration parameter and the transition rates increase. In short, the bifurcation of transition paths facilitates transition.

Finally we examine the coupling of more genes. As predicted by the theory, the bifurcation in the transition paths is not dependent on the number of genes and the actions $\Delta W$ exhibit a similar dependency on $\alpha$ (Fig. \ref{nd_result}a, upper panels). Numerical simulation reveals that bifurcated MPEP has an interesting shape. As illustrated in Fig. \ref{nd_result}b for $n=5$ and $\alpha=4$, we plot how the states of individual genes differ from the arithmetic average of all states, and it shows that only one gene is in the leading position (blue curve) and the remaining ones follow with the same states (other curves). We numerically verify that the pattern of one leading/the rest following is the optimal one and other patterns are associated with a higher action. This finding is consistent with the probabilistic view that transient alternation in the state of one gene against the drift is more probable than alternating multiple simultaneously. Numerical analysis also shows that the actions associated with non-bifurcated MPEP scale linearly with the number of genes $n$, suggesting an exponential scaling law in the transition rate. The actions for bifurcated MPEP scales sublinearly with $n$ (Fig. \ref{nd_result}c). This pinpoints that the bifurcation in MPEP leads to a softer dependency between the transition rate and the size of the system.

\section{Discussions}\label{discussion}

In this work, taking the model of coupled bistable gene circuits as an example, we consider a general model of coupled bistable systems and derive a criterion to determine whether the most probable escape paths of the transition bifurcate or not. We show that in the present setup where the steady states are not modulated by $\alpha$ and $n$, this criterion is independent on the number of individual bistable systems. Furthermore, we apply our criterion to the model of coupled bistable gene circuits and verify the theory's predictive power numerically. Numerical analysis reveals that only one bistable system takes the leading position in the bifurcated MPEP. We also show that the transition rates associated with non-bifurcated MPEP scale exponentially with the number of bistable systems while the coupled systems with bifurcated MPEP exhibit a softer scaling behavior. 

The theoretical criterion in this work is developed for a restricted class of systems satisfying the conditions that every individual system is bistable and its steady states are not affected by the coupling and the MPEP passes through the saddle of the coupled system. One may study bifurcation of a wide range of systems by following our procedures and making appropriate changes. For example, it is possible to study the coupling of identical systems where every individual system is not bistable but the coupled system is, as well as to study the coupled systems whose the steady states are dependent on the coupling. The saddle point could also be replaced by the "global maximum along the dominant path" defined by the point along the MPEP where the drift along the path changes its sign (refer to \cite{Feng2014}). Finally, one may extend the theory to study the coupling of multidimensional bistable systems, though one needs to numerically search for the optimal transition path for one bistable system and integrate $z_1$ (now a matrix rather than a scalar) along this path. In summary, our work provides with a convenient method to study the bifurcation of transition paths of coupled bistable systems and may lead to wide range of practical insights.

\section*{Acknowledgments}

C.T. and N.M. acknowledge Kim Sneppen and Erik Aurell for helpful discussions. This study is funded by the Danish National Research Foundation through the Center for Models of Life (C.T. and N.M.) and Center for Bacterial Stress Response and Persistence (N.M.).

\setcounter{equation}{0}
\renewcommand{\theequation}{A\arabic{equation}}

\section*{Appendix: Detailed Derivations in Theory}

\subsection{Non-bifurcating Path Is A Zero-Energy Path of the Freidlin-Wentzell Hamiltonian}\label{app1}

Here we show that the non-bifurcating path of a coupled bistable system is a zero-energy trajectory of the corresponding Freidlin-Wentzell Hamiltonian. First we restrict our attention to one bistable system. Since it is one-dimensional, there exists only one transition path in the zero-noise limit and it is obviously a zero-energy path of the corresponding Freidlin-Wentzell Hamiltonian $\mathcal{H}_1 = \frac{1}{2} p^2 g(x,x) + p f(x,x) = 0$ where we write the drift and variance explicitly. We then discuss the non-bifurcating path of a coupled bistable system. Note that at every point along the non-bifurcating path, $x_1=x_2=\cdots=x_n=h(\mathbf{x})$ and $p_1=p_2=\cdots=p_n$ hold, and we have that the drifts for every individual bistable systems are the same and the diagonal elements of the covariance matrix $\mathbf{G}$ are equal as well. One may then express the corresponding Freidlin-Wentzell Hamiltonian in the form that $\mathcal{H} = \frac{1}{2} \mathbf{p}^T\mathbf{G}(\mathbf{x})\mathbf{p} + \mathbf{F}^T\mathbf{p} = n\left(\frac{1}{2}p^2g(x,x) + pf(x,x)\right)$ where we use $x$ and $p$ to represent the state and momentum of every bistable system. In other words, the coupling vanishes effectively and the Hamiltonian is only dependent on the transition of every individual bistable system. Since the transition path of every individual system is zero-energy, we can claim that the non-bifurcating path of the coupled system should satisfy $\mathcal{H}=0$.

\subsection{Computation of the Hessian Matrix $\mathbf{Z}$}\label{app2}
We compute the eigenvalues of the hessian matrix $\mathbf{Z}=\nabla\nabla W$ along the non-bifurcating path. The element $z_{i,j}$ in the matrix $\mathbf{Z}$ contains information about the interaction between the $i$-th and the $j$-th bistable systems. Recall that all the bistable systems take the same state along the non-bifurcating path, and we have that all the diagonal elements should take the same value (denoted as $z_1+z_2$) and same for the off-diagonal elements (denoted as $z_2$), though we should notice that the diagonal elements may not equal to the off-diagonal ones. The hessian matrix $\mathbf{Z}$ then takes the form of $z_1\mathbf{I}_{n\times n} + z_2\mathbf{1}_{n\times n}$ and it is straightforward to show that the eigenvalues are $z_1$ (repeat $n-1$ times) and $z_1+nz_2$.

The non-bifurcating path also allows us to simplify the computation of the drifts and the noise levels in Eq. \ref{hessian_full}. Take the drifts as an example. Since all the bistable systems take the same state, for any bistable systems $i$, $j$, $k$ and $l$ ($i\neq j$, $k\neq l$), we have $x_i=x_j=x_k=x_l$ and therefore $F_i = F_j = F_k = F_l$. The linearization of the drifts is then $B_{i,j} = \partial f(x_i, h(\mathbf{x}))/\partial x_j = f_2'(x_i, h(\mathbf{x})) \partial h(\mathbf{x})/\partial x_j$. Note that the function $h$ is symmetric for all individual bistable systems, and we have $\partial h(\mathbf{x})/\partial x_j = \partial h(\mathbf{x})/\partial x_l$. We then show that $B_{i,j} = f_2'(x_k, h(\mathbf{x})) \partial h(\mathbf{x})/\partial x_l =  \partial f(x_k,h(\mathbf{x}))/\partial x_l = B_{k,l}$. In other words, the off-diagonal elements of the linearization matrix $\mathbf{B}$ take the same value. Similarly, one can prove that the diagonal elements also take the same value and we can express $\mathbf{B}$ in the form of $\mathbf{B} = f_1'\mathbf{I}_{n\times n} + f_2'h'\mathbf{1}_{n\times n}$ where the arguments to the functions are $x$ and $h'$ is defined as $\partial h(\mathbf{x})/\partial x_i$. Meanwhile, we may also simplify the hessian of the drift $\nabla\nabla F_k$. By noticing that $p_1=p_2=\cdots=p_n$, we have that
\begin{align*}
  &\sum_k p_k\nabla\nabla F_k = p(f_{11}''+nf_2'(h''_{s}-h''_{a}))\mathbf{I}_{n\times n} \\
  & + p(2f_{12}''h' + nf_{22}''h'^2 + nf_2'h''_{a})\mathbf{1}_{n\times n}
\end{align*}
where $p$ is the momentum of one bistable system, $h''_{s}$ is defined as $\partial^2 h(\mathbf{x})/\partial x_i^2$ and $h''_{a}$ is defined as $\partial^2 h(\mathbf{x})/\partial x_i\partial x_j$ for $i\neq j$. Following the same procedure, one can show that $\mathbf{C} = g_1'p\mathbf{I}_{n\times n} + g_2'h'p\mathbf{1}_{n\times n}$ and
\begin{align*}
  &\sum_k p_k^2\nabla\nabla G_{k,k} = p^2(g_{11}''+ng_2'(h''_s - h''_a))\mathbf{I}_{n\times n} \\
   &+ p^2(2g_{12}''h'+ng_{22}''h'^2 + ng_2'h''_a)\mathbf{1}_{n\times n}
\end{align*}

We are readily to obtain the eigenvalues of the hessian matrix $\mathbf{Z}$ by plugging the equations above into Eq. \ref{hessian_full}. By making use the facts that $h' = 1/n$, $h''_a = (1-\alpha)/(n^2 x)$ and $h''_s = (1-\alpha)/(n^2 x) + (\alpha-1)/nx$ along the non-bifurcating path, we have Eq. \ref{z1_dt} and

\begin{align}
  &\frac{\mathrm{d}z_2}{\mathrm{d}t} = - g(2z_1z_2+nz_2^2) - \frac{2}{n}p(z_1g_2' + nz_2g_1' + nz_2g_2')\notag\\
   & - \frac{1}{2n}p^2 (2g_{12}'' + g_{22}'')  - \frac{1-\alpha}{2nx}p^2 g_2' - \frac{1-\alpha}{nx}pf_2' \notag\\
   & - \frac{1}{n}p(2f_{12}''+f_{22}'') - \frac{2}{n}(z_1f_2'+nz_2f_1'+nz_2f_2') \label{z2_dt}
\end{align}

\subsection{Analysis of $z_1+nz_2$} \label{app3}

We analyze the eigenvalue $z_1+nz_2$ along the non-bifurcating path. By making use of Eq. \ref{z1_dt} and Eq. \ref{z2_dt}, we show that this eigenvalue follows
\begin{align}
  &\frac{\mathrm{d}(z_1+nz_2)}{\mathrm{d}t} =- g(z_1+nz_2)^2 - 2(z_1+nz_2)(g_1'+g_2')p\notag\\
   &  - p(f_{11}''+2f_{12}''+f_{22}'') - \frac{1}{2}p^2(g_{11}''+2g_{12}''+g_{22}'') \notag\\
   & - 2(z_1+nz_2)(f_1'+f_2') \label{z1nz2_dt}
\end{align}
In Eq. \ref{z1nz2_dt}, the term $g_1'+g_2'$ is equivalent to $\mathrm{d}g(x,x)/\mathrm{d}x$, $g_{11}''+2g_{12}''+g_{22}''$ is equivalent to $\mathrm{d}^2 g(x,x)/\mathrm{d}x^2$ and similar for the terms of $f$. Since $f(x,x)$ and $g(x,x)$ are the drift and noise induced by one bistable system, we claim that the eigenvalue $z_1+nz_2$ corresponds to the direction of perturbation where the coupling of multiple bistable systems effectively vanishes. The only possible direction is the one along the non-bifurcating path. Therefore, the eigenvalue $z_1+nz_2$ should not be considered in our context as it does not break the non-bifurcating assumption. We should then focus on the eigenvalue $z_1$, as discussed in Section \ref{theory}.

\subsection{Validation of Low-Noise Limit in Gillespie Simulation}\label{verification_FW}

We verify that the Gillespie simulation in Fig. \ref{model}b is carried out in the low-noise limit. We perform simulation with three volumes ($V=30, 40, 45$) and estimate the actions by fitting the simulation data to $T=C\mathrm{e}^{\Delta W/\epsilon}$ where $T$ is the mean transition time and $\epsilon=1/V$. We show that the estimated actions are in good agreement with the values associated with the MPEP which are computed in the low-noise limit (Fig. \ref{nd_result}, Table \ref{comparison_dW}). Furthermore, we plot the MPEP on top of the heat plots in Fig. \ref{model}b and we show that the transition paths sampled from Gillespie simulation match the MPEP. Therefore, we claim that our Gillespie simulation is in the low-noise limit.

\begin{table}
  \caption{\label{comparison_dW} Estimation of the actions ($\Delta W$) from Gillespie simulation.} 
  \setlength\tabcolsep{10pt}
  \begin{tabular}{l l l l l l}
    \hline
    \multicolumn{2}{l}{} & $\alpha=1$ & $\alpha=2$ & $\alpha=4$ & $\alpha=7.5$ \\
    \hline
    \multirow{3}{*}{Mean Time} & $V=30$ & $1.37*10^6$ & $6.02*10^5$ & $1.15*10^5$ & $1.99*10^4$ \\
    & $V=40$ & $5.44*10^7$ & $2.54*10^7$ & $2.51*10^6$ & $2.55*10^5$ \\
    & $V=45$ & $2.97*10^8$ & $1.25*10^8$ & $1.31*10^7$ & $8.02*10^5$ \\
    \hline
    \multicolumn{2}{l}{Estimated $\Delta W$ (Gillespie)} & 0.360 & 0.358 & 0.315 & 0.248 \\
    \hline
    \multicolumn{2}{l}{Estimated $\Delta W$ (MPEP)} & 0.360 & 0.360 & 0.325 & 0.259 \\
    \hline
  \end{tabular}
\end{table}


\begin{thebibliography}{5}
\bibitem{Ferrell2001} J.E. Ferrell, W. Xiong, Chaos. {\bf 11}, 227 (2001).
\bibitem{Ferrell2002} J.E. Ferrell, Curr. Opin. Cell. Biol. {\bf 14}, 140 (2002).
\bibitem{Veening2008} J.W. Veening, W.K. Smits, O.P. Kuipers, Annu. Rev. Micro. {\bf 62}, 193 (2008).
\bibitem{Balazsi2011} G. Balazsi, A. van Oudenaarden, J.J. Collins, Cell. {\bf 144}, 910 (2011)
\bibitem{Aurell2002} E. Aurell, K. Sneppen, Phys. Rev. Lett. {\bf 88}, 048101 (2002)
\bibitem{Wang2011} J. Wang, K. Zhang, L. Xu, E. Wang, Proc. Natl. Acad. Sci. U.S.A. {\bf 108}, 8257 (2011)
\bibitem{Miller2001} M.B. Miller, B.L. Bassler, Annu. Rev. Microbiol. {\bf 55}, 165 (2001)
\bibitem{Muller2006} J. Muller, C. Kuttler, B.A. Hense, M. Rothballer, A. Hartmann, J. Math. Biol. {\bf 53}, 672 (2006)
\bibitem{Maisonneuve2014} E. Maisonneuve, K. Gerdes, Cell {\bf 157}, 539 (2014)
\bibitem{Cataudella2012} I. Cataudella, A. Trusina, K. Sneppen, K. Gerdes, N. Mitarai, Nuc. Acids Res. {\bf 40}, 6424 (2012).
\bibitem{Fasani2013} R.A. Fasani, M.A. Savageau, Proc. Natl. Acad. Sci. U.S.A. {\bf 110}, E2528 (2013).
\bibitem{Maier1996} R.S. Maier, D.L. Stein, J. Stat. Phys. {\bf 83}, 291 (1996)
\bibitem{Gillespie1977} D.T. Gillespie, J. Phys. Chem. {\bf 81}, 2340 (1977).
\bibitem{Note} If we expand the chemical master equation for the model of coupled bistable gene circuits using Kramers-Moyal expansion and preserve up to the second order, we obtain a Fokker-Planck equation (Eq. \ref{Fokker-Planck}) with a drift vector equal to the production rates minus the degradation rates and a diagonal diffusion matrix where the diagonal elements equal to the sum of the production and degradation rates. By defining the Langevin equation using Eq. \ref{fuv} and \ref{guv}, we will obtain the same Fokker-Planck equation and thus the definition is physical. Alternatively, one may derive chemical Langevin equation from the master equation and it contains the sum of two noise terms: one for production $\sqrt{\epsilon}\sqrt{k_0+k_1v^4/(v^4+S^4)}\xi_i$ and one for degradation $\sqrt{\epsilon}\sqrt{\gamma u}\tilde\xi_i$ \cite{Gillespie2000}. Within the scope of this manuscript, these two formulations are equivalent as they give the same Fokker-Planck equation.
\bibitem{Maier1993} R.S. Maier, D.L. Stein, Phys. Rev. E. {\bf 48}, 931 (1993)
\bibitem{Freidlin2012} M.I. Freidlin, A.D. Wentzell, Random Perturbations of Dynamical Systems, 3rd Ed. (Springer-Verlag, Berlin, 2012)
\bibitem{Maier1997} R.S. Maier, D.L. Stein, SIAM J. Appl. Math. {\bf 57}, 752 (1997)
\bibitem{Risken1996} H. Risken, The Fokker-Planck Equation: Methods of Solution and Applications, 2nd Ed. (Springer-Verlag, Berlin, 1996)
\bibitem{Naeh1990} T. Naeh, M.M. Klosek, B.J. Matkowsky, Z. Schuss, SIAM J. Appl. Math. {\bf 50}, 595 (1990)
\bibitem{Kupferman1992} R. Kupferman, M. Kaiser, Z. Schuss, E. Ben-Jacob, Phys. Rev. A. {\bf 45}, 745 (1992)
\bibitem{Note3} Eq. \ref{FW_Hamiltonian} corresponds to the system constructed by preserving the Kramers-Moyal expansion of the chemical master equation up to the second-order term (cf. \cite{Note}).
\bibitem{Heymann2008PRL} M. Heymann, E. Vanden-Eijnden, Phys. Rev. Lett. {\bf 100}, 140601 (2008).
\bibitem{Heymann2008CPAM} M. Heymann, E. Vanden-Eijnden, Commun. Pure Appl. Math. {\bf 61}, 1052 (2008).
\bibitem{Note2} The numerical simulation is carried out with Eq. \ref{FW_Hamiltonian} which is a quadratic approximation to $\mathcal{H}_{full}=\sum_i \left[\left(k_0+k_1\frac{\bar{x}^4}{\bar{x}^4+S^4}\right)(\mathrm{e}^{p_i}-1)+\gamma x_i(\mathrm{e}^{-p_i}-1)\right]$ where $p_i$ is the conjugate momentum for $x_i$ \cite{Freidlin2012}. To verify the validity of the quadratic approximation, we perform numerical simulation with $\mathcal{H}_{full}$. With the present numerical accuracy, we do not observe any difference in the MPEP and the associated actions $\Delta W$ differ by less than 2\% for the two regimes. All the conclusions presented in Section \ref{application} remain valid for the non-approximating regime. We therefore claim that our work based on Eq. \ref{FW_Hamiltonian} is a valid analysis for the chemical master equation.
\bibitem{Gillespie2000} D.T. Gillespie, J. Chem. Phys. {\bf 113}, 297 (2000).
\bibitem{Feng2014} H. Feng, K. Zhang, J. Wang, Chem. Sci. {\bf 5}, 3761 (2014).
\end{thebibliography}
\end{document}